# Effects of Quantum and Dielectric Confinement on the Emission of Cs-Pb-Br Composites

Sebastián Caicedo-Dávila[1,a,*], Pietro Caprioglio[2,3], Frederike Lehmann[1], Sergiu Levcenco[1], Martin Stolterfoht[2], Dieter Neher[2], Leeor Kronik,[4] and Daniel Abou-Ras[1]

[1]Helmholtz-Zentrum Berlin für Materialien und Energien, Hahn-Meitner-Platz 1. 14109 Berlin, Germany
[2]Institute of Physics and Astronomy, University of Potsdam, 14476 Potsdam, Germany
[3]Department of Physics, University of Oxford, Clarendon Laboratory, Parks Road, Oxford, UK
[4]Department of Molecular Chemistry and Materials Science, Weizmann Institute of Science, Rehovoth 76100, Israel
[*]sebastian.caicedo@tum.de

## Abstract

The halide perovskite $CsPbBr_3$ belongs to the Cs-Pb-Br material system, which features two additional thermodynamically stable ternary phases, $Cs_4PbBr_6$ and $CsPb_2Br_5$. The coexistence of these phases and their reportedly similar photoluminescence have resulted in a debate on the nature of the emission in these systems. Here, we combine optical and microscopic characterization with an effective mass, correlated electron-hole model of excitons in confined systems, to investigate the emission properties of the ternary phases in the Cs-Pb-Br system. We find that all Cs-Pb-Br phases exhibit green emission and the non-perovskite phases exhibit photoluminescence quantum yields orders of magnitude larger than $CsPbBr_3$. In particular, we measure blue- and red-shifted emission for the Cs- and Pb-rich phases, respectively, stemming from embedded $CsPbBr_3$ nanocrystals. Our model reveals that the difference in emission shift is caused by the combined effects of nanocrystal size and different band mismatch. Furthermore, we demonstrate the importance of including the dielectric mismatch in the calculation of the emission energy for Cs-Pb-Br composites. Our results explain the reportedly limited blue shift in $CsPbBr_3@Cs_4PbBr_6$ composites and rationalize some of its differences with $CsPb_2Br_5$.

---

[a] Present Address: Physics Department, TUM School of Natural Sciences
Technical University of Munich, 85748 Garching

# 1. Introduction

The impressive progress in power conversion efficiency of solar cells based on halide perovskite-type (HaPs) compounds, from ~4% in 2009[1] to more than 25% in recent years[2] has prompted significant research interest in these materials, not only for photovoltaics, but also for other optoelectronic applications, such as light emission and detection.[3–6] $CsPbBr_3$ is a prototypical model HaP,[7] which belongs to the Cs-Pb-Br material system with a great potential for applications in light-emitting diodes[8–14] and lasers.[15–17] Thus, understanding the emission properties of Cs-Pb-Br compounds is fundamental for the design of better materials and devices. The Cs-Pb-Br material system includes two additional, non-perovskite-type, ternary phases: The Cs-rich $Cs_4PbBr_6$ and the Pb-rich $CsPb_2Br_5$, sometimes referred to as the zero-dimensional (0D) and two-dimensional (2D) phases, respectively, owing to their crystal structure (*cf.* Figure S1). Both non-perovskite-type phases are thermodynamically stable and have been synthesized as single crystals, thin films, and nanocrystals.[18–23]

The coexistence of Cs-Pb-Br phases and its influence on the optoelectronic properties is well established.[24–27] In recent years a strong green luminescence has been measured in both $CsPb_2Br_5$ and $Cs_4PbBr_6$, which appears to be in contrast with their large band-gap energies, ~3.8 eV to 4.0 eV.[10,18,28–34] This motivated a debate on whether such luminescence is caused by intrinsic factors[35–43] or by nanocrystals (NCs) of the perovskite-type phase, $CsPbBr_3$ (with a band-gap energy in the green region of the visible spectrum, ~2.4 eV),[7,42,44–54] embedded in the non-perovskite matrix. Although substantial evidence in favor of the embedded NCs hypothesis have been reported,[49,50,52–57] open questions remain, notably the exact mechanism that enhances the luminescence and the limited blue shift measured with decreasing NC size.[56]

In the present work, we combine photoluminescence (PL) and energy-dispersive X-ray (EDX) spectroscopies, as well as cathodoluminescence (CL) hyperspectral imaging and theoretical modeling to investigate the mechanisms behind the green emission in all three ternary phases of the Cs-Pb-Br material system. Our results provide additional evidence for the green emission in $Cs_4PbBr_6$ and $CsPb_2Br_5$ stemming from embedded $CsPbBr_3$ NCs. Furthermore, we model the exciton emission in $CsPbBr_3@Cs_4PbBr_6$ and $CsPbBr_3@CsPb_2Br_5$ composites using an effective mass model, which includes a simplified treatment of electron-hole correlation, as well as the band gap and dielectric mismatch at the interfaces. Comparing this model with the commonly used effective mass model[58,59] suggests that accounting for the effects of finite confinement potentials and the Coulomb interactions with image charges at the interface can result in better estimates of NC sizes. The model explains the limited blue shift in small $CsPbBr_3$ NCs embedded in $Cs_4PbBr_6$. Furthermore, differences in the type of confinement of $CsPbBr_3$ NCs in $Cs_4PbBr_6$ and $CsPb_2Br_5$ explain the stronger emission of the former case and the measured red shift of the latter case.

# 2. Results and Discussion

Powder samples of the Cs-Pb-Br ternary phases – $CsPbBr_3$, $CsPb_2Br_5$ and $Cs_4PbBr_6$ – were synthesized as described in Section S.1 of the Supporting Information (SI). The three ternary phases exhibit PL emissions in the range from 2.3 to 2.5 eV, as shown in Figure 1. The integrated PL intensities of the $CsPb_2Br_5$ and $Cs_4PbBr_6$ are about one and three orders of magnitude larger, respectively, than that of the $CsPbBr_3$. Commensurately, the PL quantum yield (PLQY) at 1 sun equivalent illumination is $3 \times 10^{-3}\%$, 5%, and $8 \times 10^{-4}\%$ for $CsPb_2Br_5$, $Cs_4PbBr_6$, and $CsPbBr_3$, respectively. It is also noteworthy that the PL peak of green luminescence in the non-perovskite-type phases is shifted with respect to that of $CsPbBr_3$. A blue shift of 50 meV and a red shift of 40 meV were measured for the PL peaks of $Cs_4PbBr_6$ and $CsPb_2Br_5$. The emission shifts are consistent among spectra measured on different sample areas and are significant, considering our measurement resolution (~5 meV, see Section S.2 of the SI) and the thermal energy at room temperature (~25 meV).

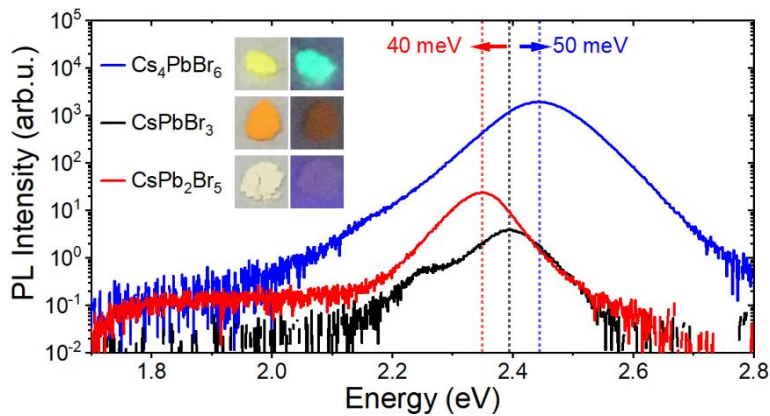

**Figure 1.** PL spectra of the three Cs-Pb-Br ternary phases. All the phases exhibit green luminescence between 2.3 and 2.5 eV. Spectra are plotted in semi-logarithmic scale to show the differences of the PL intensity. Insets show photographs of the powder samples under ambient light (left) and 205 nm LED illumination (right).

We investigated the origin of the green emission in the $CsPb_2Br_5$ and $Cs_4PbBr_6$ samples using scanning electron microscopy (SEM), CL imaging, and EDX spectroscopy – see Figures 2 to 4 below. Experimental details are given in Section S.2 of the SI. The CL maps for both $Cs_4PbBr_6$ and $CsPb_2Br_5$ show that luminescence in the visible spectral range stems from localized emitters, embedded in a solid matrix. Panchromatic (spectral range from 1.95 eV to 2.8 eV) and 500 nm bandpass-filtered maps are identical, confirming that the materials exhibit green emission (peak emission between 2.3 to 2.5 eV, *cf.* Figure 2 and 3). EDX elemental maps show that for $Cs_4PbBr_6$ the emitters exhibit higher Pb and lower Cs and Br counts (see Figure 2c-f), whereas for $CsPb_2Br_5$ the emitters exhibit lower Pb and higher Cs counts, but Br counts remain unchanged (*cf.* Figure 3c-e). The compositions of the emitting regions relative to the matrix suggest that they are $CsPbBr_3$. Therefore, we conclude that the green emitters are $CsPbBr_3$ NCs embedded in the host $Cs_4PbBr_6$ or $CsPb_2Br_5$ matrix.

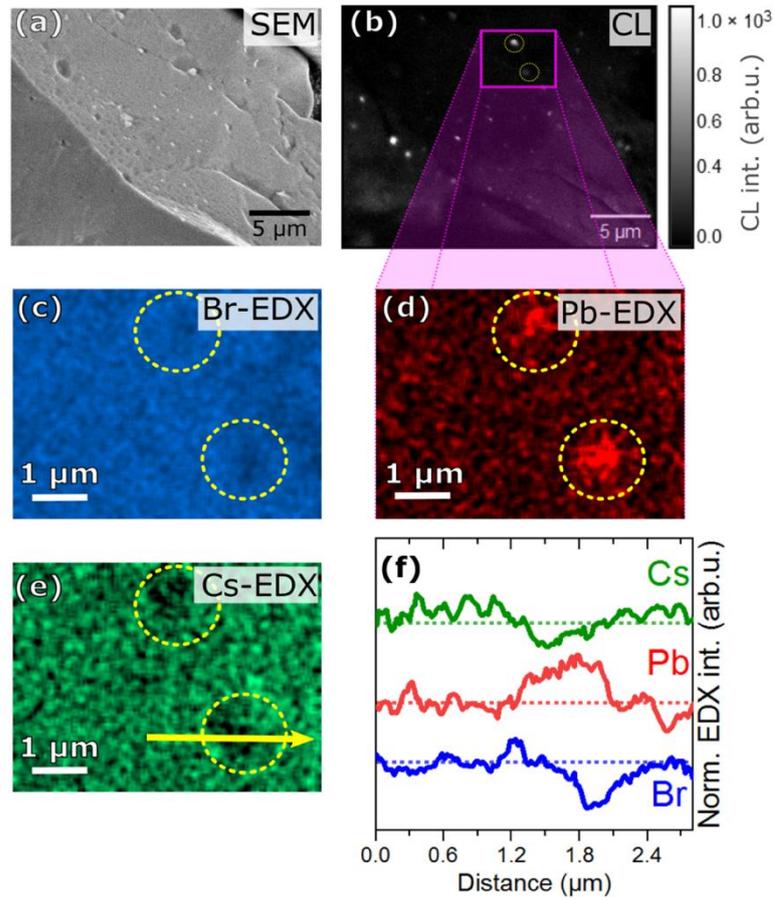

**Figure 2.** (a) SEM image of Cs$_4$PbBr$_6$ crystal and (b) a corresponding CL intensity map, filtered at 500 nm, showing the localized emitters embedded in the crystal. (c-e) EDX elemental distribution maps and (f) normalized line-scans — along the yellow arrow in (e) — of a magnified region, showing clear enrichment of Pb as well as depletion of Cs and Br, correlated with the yellow-circled position of highly luminescent clusters.

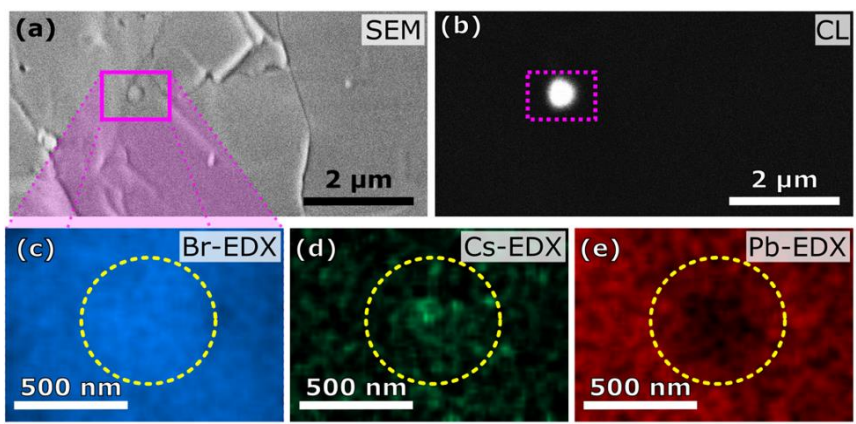

**Figure 3.** (a) SEM image of a CsPb$_2$Br$_5$ crystal and (b) a corresponding CL intensity map, filtered at 500 nm. EDX elemental distribution maps of (c) Br, (d) Cs, and (e) Pb show a clear enrichment

of Cs and depletion of Pb, correlated with the yellow-circled position of highly luminescent clusters.

In order to investigate the spectral features of the CsPbBr$_3$ NCs, we performed hyperspectral CL mapping, using a low acceleration voltage (3.5 kV) in order to avoid beam damage and reduce the interaction volume. This allows us to improve the spatial resolution of the CL signal around the NCs. The CL map of Cs$_4$PbBr$_6$ acquired at larger magnification shows that luminescence stems from clusters of NCs, rather than individual emitters (see Figure 4a). The NCs exhibit emission maxima in the range from 2.44 to 2.47 eV (spectrum 1 in Figure 4b), which agrees well with the PL measurement (*cf.* Figure 1). Furthermore, no emission was detected from the matrix material (spectrum 2 in Figure 4b). For the CsPb$_2$Br$_5$ sample, the CL spectra of the emitters peak between 2.35 and 2.37 eV (spectrum 1 in Figure 4e), in good agreement with the PL results (*cf.* Figure 1). The matrix material exhibits no sharp CL peak. However, we detected a wide band of weak emission centered at ~1.9 eV (spectrum 2 in Figure 4e) when measuring CL on matrix regions. This observation is consistent with defect emission at the interface, we measured for CsPb$_2$Br$_5$/CsPbBr$_3$ films in previous work (*cf.* Ref. [60] and its supplemental material).

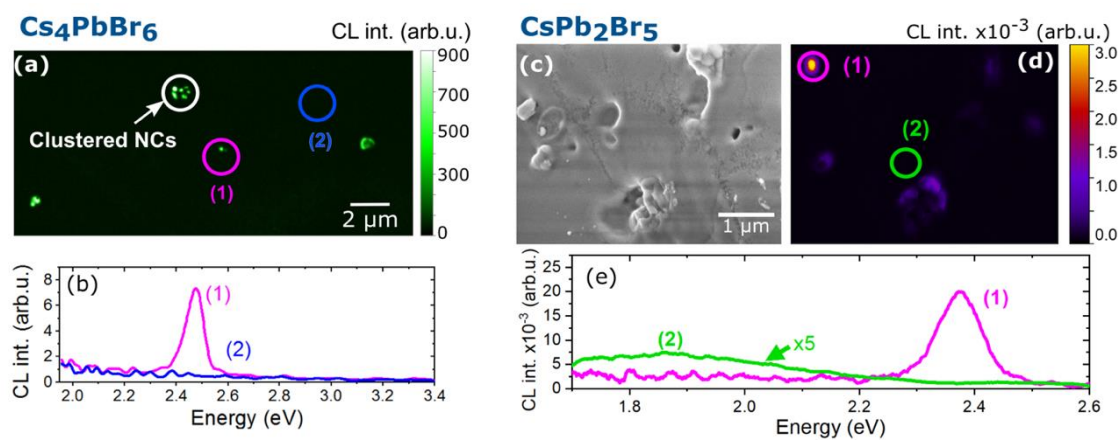

**Figure 4.** (a) CL intensity map, filtered at 500 nm, on a flat surface of a Cs$_4$PbBr$_6$ crystal. (b) Corresponding CL spectra (hyperspectral measurement without filter) of selected areas. (c) SEM image of the surface of CsPb$_2$Br$_5$. (d) Corresponding CL map, filtered at 550 nm, and (e) CL spectra of selected areas. Spectrum 2 is magnified x5 for visibility.

Our CL-EDX correlative characterization clarifies the origin of the green emission in Cs$_4$PbBr$_6$ and CsPb$_2$Br$_5$. In order to understand the recombination mechanism and the nature of emission, we performed intensity-dependent PLQY measurements by exciting only the perovskite phase using a 445 nm laser (see Figure 5). Because of the excitation wavelength used, we can exclude charge transfer from the wide-gap, non-perovskite phases to the narrower perovskite ones. By investigating the slope $\gamma$ of the dependence of the emitted photon flux ($I_{PL} = I^\gamma$) and the slope $k$ of the dependence of the PLQY (PLQY $= I^k$) on the illumination intensity ($I$) in a log-log plot, it is possible to access the charge recombination mechanisms. The green emission from the Cs$_4$PbBr$_6$ phase exhibits $\gamma = 1$ and $k \approx 0$. This indicates an excitonic character because it is known that a value of $\gamma = 1$ and intensity-independent PLQY correspond to radiative recombination of free excitons.[61–63] In contrast, the values of $\gamma = 1.5$ and $k = 0.5$ for CsPb$_2$Br$_5$ indicate a deviation from pure excitonic emission. This suggests that the green emission in this phase is dominated by nonradiative recombination. For the CsPbBr$_3$ phase, we find $\gamma = 1.1$ and $k = 0.1$, slightly deviating from a pure excitonic emission, and the absolute PLQY values are lower compared with those of Cs$_4$PbBr$_6$. This behavior suggests a stronger contribution of nonradiative recombination processes occurring in bulk CsPbBr$_3$, as compared with NC domains in Cs$_4$PbBr$_6$, suggesting that the latter is less defective. The contribution of nonradiative recombination processes can also be associated to the low PLQY of out Cs-Pb-Br composites, compared with the previous reports.[31,32,64]

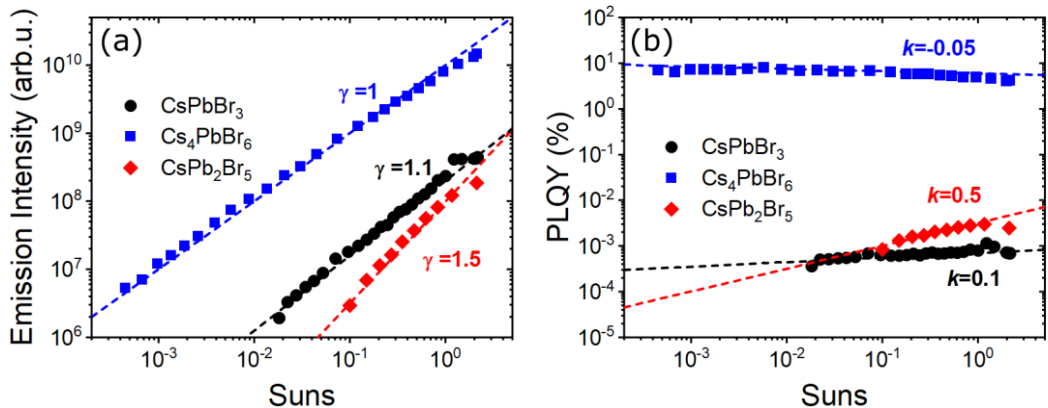

**Figure 5.** (a) Photoluminescence emitted photon flux and (b) PLQY as a function of the excitation intensity. The data are plotted in a logarithmic scale and linearly fit. The slope of the PL ($\gamma$) and PLQY ($k$) curves reveals the type of dominant recombination, see text for details.

The microscopy and spectroscopy results confirm consistently that the non-perovskite-type Cs-Pb-Br samples contain CsPbBr$_3$ NCs embedded in a solid matrix. The confinement of the NCs also explain the blue shift of the PL measured for Cs$_4$PbBr$_6$, with respect to bulk CsPbBr$_3$ emission (Figure 1). The best achievable spatial resolution for our CL experiments is in the range of ~60-80 nm, as estimated from the density of Cs$_4$PbBr$_6$ (4.29 g/cm$^3$)[65,66] and CsPb$_2$Br$_5$ (5.76 g/cm$^3$),[34,67] the electron beam parameters, and Gruen's equation (Eq. S1 in Section S.2 of the SI). This resolution limit, as well as the diffusion of charge carriers before they recombine, limits the accuracy of the estimated NC size directly from CL experiments. Because the magnitude of quantum and dielectric confinement effects is generally related to the size of the NCs, it is interesting to examine whether a theoretical model can be used to provide an explanation for the origin of different blue- and red-shift effects and their dependence on the size of the emitting CsPbBr$_3$ NCs.

An important and commonly used approximation to the emission in such systems, within the effective mass approximation, has been proposed by Brus in the 80s.[58,59,68] This model considers a spherical NC in an infinite potential well, where the Coulomb interaction is strongly screened, such that the exciton (electron-hole) wave function is uncorrelated. However, as stated by Brus in his original paper,[58] this model can be a poor approximation for large band gap materials with moderate NC sizes, because in those systems the Coulomb energy is comparable to the confinement energy and the electron-hole correlation can be important. Considering that CsPbBr$_3$ is indeed a (relatively) large gap material with a (relatively) significant exciton binding energy (35—60 meV),[39,69–72] we opt for a simple model, enforcing confinement in one dimension, that includes some form of electron-hole correlation, while remaining mathematically tractable and computationally inexpensive. We show below that this is sufficient for a qualitative explanation of our results.

We apply an effective mass model, proposed by Rajadell et al. for CdSe nanostructures,[73–77] The model explicitly treats the correlated electron-hole pair, including Coulomb interactions and dielectric effects, based on the method of image charges that approximate polarization terms in the exciton Hamiltonian (see Section S.3 in the SI for further details). Within this framework, the Hamiltonian that describes the electron-hole pair is given by:

$$H(\mathbf{r}_e, \mathbf{r}_h) = H^{(e)}(\mathbf{r}_e) + H^{(h)}(\mathbf{r}_h) + V_c(\mathbf{r}_e, \mathbf{r}_h) \tag{1}$$

where $H^{(i)}(\mathbf{r}_i)$ is an effective-mass, single-particle Hamiltonian, that contains image charge effects, which is described in detail in Eq. S2 of the SI. $V_c(\mathbf{r}_e, \mathbf{r}_h)$ is a generalized electron-hole Coulomb interaction potential, calculated using the method of image charges. This term describes the interaction of a charged particle with interface image charges induced by the other particle and can be written in one dimension as:[77]

$$V_c(\mathbf{r}_e, \mathbf{r}_h) = \sum_{n=-\infty}^{\infty} \frac{q_n e^2}{\varepsilon_1 \sqrt{\|\mathbf{r}_{\|,e} - \mathbf{r}_{\|,h}\|^2 + [z_e - (-1)^n z_h + nL]^2}} \tag{2}$$

where $\mathbf{r}_{\|,i}$ is the in-plane particle position (perpendicular to the confinement direction), $z_i$ is the particle position in the confined direction, $L$ is the width of the well (NC size), and $q_n = \left(\frac{\varepsilon_1 - \varepsilon_2}{\varepsilon_1 + \varepsilon_2}\right)^{|n|}$, where $\varepsilon_1$ and $\varepsilon_2$ are the dielectric constant of the NC and host, respectively. We seek a two-particle wave-function solution to this Hamiltonian, namely

$$H(\mathbf{r}_e, \mathbf{r}_h)\Psi(\mathbf{r}_e, \mathbf{r}_h) = E_{exc}\Psi(\mathbf{r}_e, \mathbf{r}_h), \tag{3}$$

where the wave function is approximated by[76]

$$\Psi(\mathbf{r}_e, \mathbf{r}_h) = N\psi_e(\mathbf{r}_e)\psi_h(\mathbf{r}_h)e^{-a\sqrt{\|\mathbf{r}_{\|,e} - \mathbf{r}_{\|,h}\|^2}}, \tag{4}$$

where $\psi_{e/h}$ are the single-particle wave functions for electron and hole, and $a$ is a variational parameter. Solving Eqs. (1)-(4) requires electron and hole effective masses as well as dielectric constant values. For the former, we use the values reported by Protesescu et al.[71] for CsPbBr$_3$. For the latter, we used density functional perturbation theory (DFPT) to obtain computed values for all three Cs-Pb-Br phases (see Section S.4 in the SI for details). All parameters used in the model are summarized in Table S1 of the SI.

When $\varepsilon_1 > \varepsilon_2$, i.e., the dielectric constant of the NC is larger than that of the host, there is dielectric confinement. The image charge induced at the interface exhibits the same sign as the confined charge and the electric field owing to the confined charge penetrates the matrix region, as shown schematically in Figure 6a. This reduces the effective dielectric constant, with respect to $\varepsilon_1$, and enhances the Coulomb interaction between electron and hole.[73,78] The electron-hole Coulomb interaction is further modulated by the overlap between the electron and hole wave functions, which depends on the degree of confinement. Consequently, the exciton binding energy will be a function not only of $\varepsilon_1$ and $\varepsilon_2$, but also of the NC size.

We first use the above model to investigate the effect of a dielectric mismatch for a NC confined by an infinite potential. We calculate the shift of the emission energy, $E_{em}$ with respect to the bulk band gap of CsPbBr$_3$, $E_g$, as a function of the NC size $L$, for various values of $\varepsilon_2$ (see Figure 6b). For small $L$, quantum confinement is dominant and the effect of dielectric mismatch on $E_{em} - E_g$ is negligible. As $L$ increases and quantum confinement is reduced, the effect of dielectric mismatch becomes more important, and $E_{em} - E_g$ decreases with $\varepsilon_2$ owing to a stronger dielectric confinement.

Dielectric effects also directly impact the exciton binding energy, $E_b$, namely the difference between the emission energy and the single-particle bandgap, as shown in Figure 6c,d. Because dielectric confinement reduces the effective dielectric constant and Coulomb screening, it increases $E_b$. Specifically, $E_b$ rapidly decreases with $L$ for $\varepsilon_2 < \varepsilon_1$. This dependence weakens as $\varepsilon_2$ increases, demonstrating that dielectric confinement is stronger for smaller NCs, in good agreement with previous reports.[73,77,79–81]

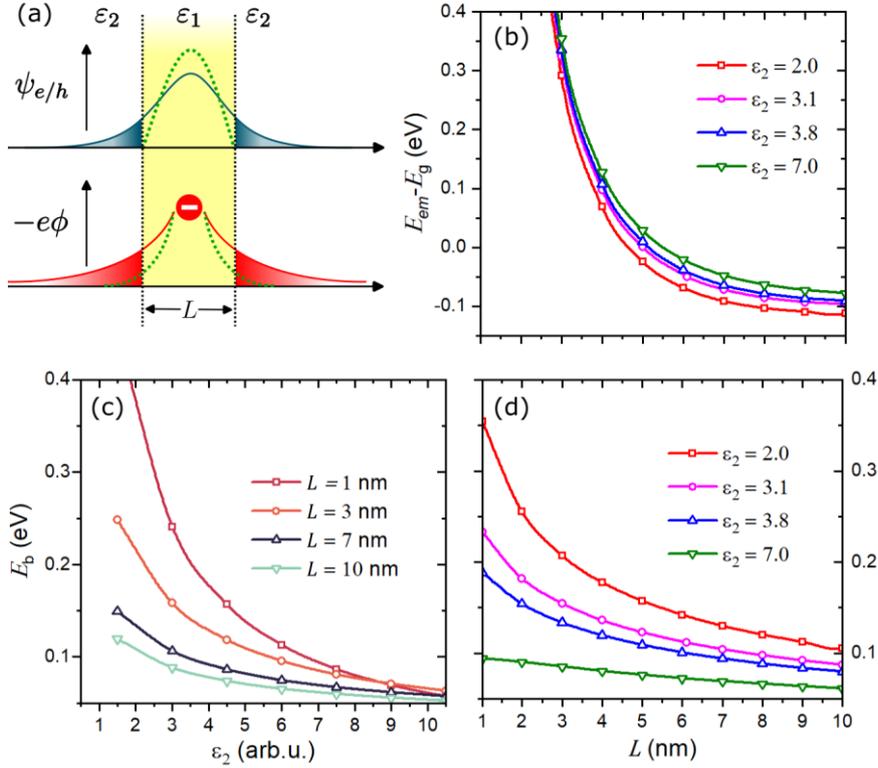

**Figure 6.** Effect of confinement and dielectric mismatch in a simple 1D model of a CsPbBr$_3$ NC. (a) Top: schematic representation of a single-particle wave function, $\psi_{e/h}$, confined in an infinite (dashed) and finite (solid) potential. Bottom: schematic representation of the effect of dielectric mismatch on the electrostatic potential, $\phi$, for $\varepsilon_1 > \varepsilon_2$ (solid) and $\varepsilon_1 < \varepsilon_2$ (dashed). (b) Calculated shift of the emission energy with respect to the bulk band gap energy, $E_{em}$ - $E_g$, as a function of the NC size for different values of $\varepsilon_2$. (c) exciton binding energy, $E_b$, as a function of $\varepsilon_2$ for various NC sizes. (d) $E_b$ as a function of $L$ for various $\varepsilon_2$.

Next, we investigate the effect of a finite confinement potential on the emission energy, by including the band mismatch between NC and matrix material, i.e., Cs$_4$PbBr$_6$ or CsPb$_2$Br$_5$ (see Section S.3 in the SI for details). We use the literature-reported band-gap energies of the Cs-Pb-Br phases, i.e., 2.4 eV for CsPbBr$_3$ (in reasonable agreement with our measurements),[7,44–48] 4.0 eV for Cs$_4$PbBr$_6$[28–32] and 3.7 eV for CsPb$_2$Br$_5$.[18,33,34] The band mismatch for CsPbBr$_3$/Cs$_4$PbBr$_6$ results in a type I band alignment, i.e., both hole and electron wave functions are confined, as deduced from both theoretical calculations[82] and (X-ray and UV) photoelectron spectroscopy.[83,84] We model the type I alignment using a fixed band mismatch $\frac{\Delta E_g}{2}$ (where $\Delta E_g = E_g^{Cs_4PbBr_6} - E_g^{CsPbBr_3}$) for the conduction and valence band mismatch. In the case of CsPbBr$_3$/CsPb$_2$Br$_5$, some theory reports a quasi-type I alignment (only one carrier confined),[85] while experimental characterization assigns a type II band alignment,[86] i.e., the electrons and holes are confined to different regions, which reduces their wave function overlap. We modeled this by setting a small but negative valence band offset $\Delta E_V = -0.1$ eV and a corresponding conductions band offset of $\Delta E_C = \Delta E_g + 0.1$ eV (where $\Delta E_g = E_g^{CsPb_2Br_5} - E_g^{CsPbBr_3}$).

The effect of the finite band mismatch is apparent in the calculated shift in emission energy $E_{em} - E_g$, shown in Figure 7, where the finite potential reduces the magnitude of the quantum confinement effect. This effect is stronger in $Cs_4PbBr_6$ than it is in $CsPb_2Br_5$, partly because of the larger band gap of the former, but mostly due to the more effective quantum confinement of the type I alignment. Generally, for a finite confinement potential the wave function extends into the barrier regions (*cf.* Figure 6a), making quantum confinement less effective and also reducing the effective dielectric constant of the system, thereby favoring dielectric confinement.[73] At the same time, the decrease in electron-hole overlap reduces the Coulomb interaction, hindering dielectric confinement. This competition results in a complex dependence of the exciton binding energy on the finite potential barriers. In the present calculation, confinement effects are apparent for $L < 6.3$ nm ($< 7$ nm) for $Cs_4PbBr_6$ ($CsPb_2Br_5$), providing a guide to the size regimes for which quantum or dielectric confinement is dominant. While the similarity to the exciton Bohr radius of Bulk $CsPbBr_3$ (~7 nm)[71] is notable, we emphasize that this is a 1D model, which should not be compared fully quantitatively to experiment.

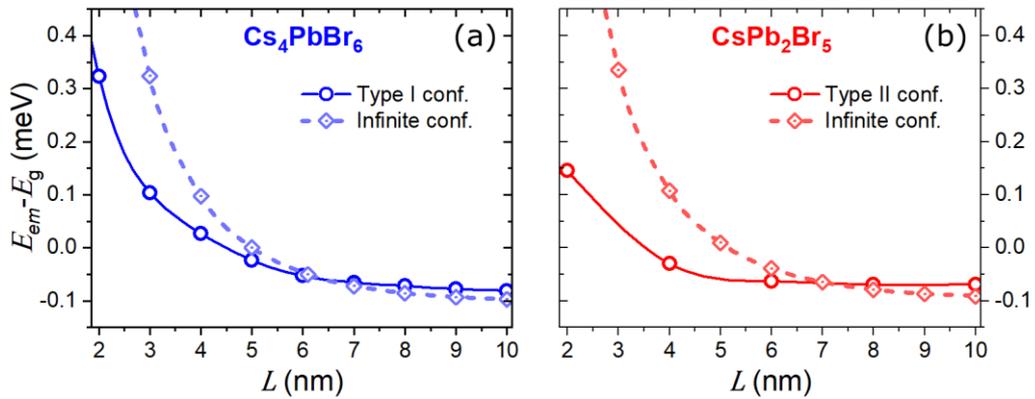

**Figure 7.** Emission energy shift, $E_{em} - E_g$, of a $CsPbBr_3$ NC embedded in (a) $Cs_4PbBr_6$, or (b) $CsPb_2Br_5$, as a function of the NC size. Dielectric mismatch is modeled with $\varepsilon_1 = 4.2$ for $CsPbBr_3$, $\varepsilon_2 = 3.1$ for $Cs_4PbBr_6$, and $\varepsilon_2 = 3.8$ for $CsPb_2Br_5$. Results for infinite (dashed curves) and finite (solid curves) confinement potentials, with type I and type II alignments for $Cs_4PbBr_6$ and $CsPb_2Br_5$, respectively, are compared.

We can now qualitatively interpret our experimental data, in light of the computed result from the above model. In the case of $Cs_4PbBr_6$, it is clear that the size of the embedded $CsPbBr_3$ NCs lies in the strong confinement regime, i.e., quantum confinement is dominant and results in the above-reported blue shift and PLQY enhancement of the exciton emission. The blue shift in this system is reduced with respect to the conventional estimations using a simple effective mass model, owing to the combined effect of a finite confinement potential and the reduced effective dielectric constant. This rationalizes the limited emission shift achieved experimentally for $CsPbBr_3@Cs_4PbBr_6$ composites.[32,56] For $CsPb_2Br_5$, a plausible explanation for the red shift measured in our PL and CL experiments is that there is a weaker quantum confinement caused by type II band alignment, coupled with larger NC sizes, expected from the similar formation enthalpies of the $CsPbBr_3$ and $CsPb_2Br_5$ phases.[26,27,60] For larger NCs, the dielectric mismatch can cause enough of an increase in the exciton binding energy to overcome the effect of the weak quantum confinement, resulting in a net red shift in $E_{em}$ with respect to the bulk $E_g$. This also can explain the large discrepancy between NC sizes estimated with a simple effective mass model and the TEM images, reported for $CsPb_2Br_5$. For this case, the error in the estimation of the NC size was found to be as large as 4 nm, which is much larger than the error for $Cs_4PbBr_6$ (0.6 nm).[87]

While the simple model provides a plausible qualitative explanation for all emission energy patterns observed in our experiments, we caution that further refinement is needed to gain a complete quantitative understanding of the PL results from the composite Cs-Pb-Br materials. First, the model used in the present work is one-dimensional, whereas the effects of dielectric and quantum confinement could be stronger and have a more complex relationship in a system confined in all three spatial dimensions. For example, the dielectric tensor of $CsPb_2Br_5$ is highly anisotropic, i.e., the effect of dielectric mismatch would depend strongly on the confinement direction. Also, in our model we did not consider changes in the electron and hole effective masses, which can also influence the exciton levels.[68,88] Furthermore, considering the exciton fine structure of the $CsPbBr_3$ NCs, including effective mass non-parabolicity using an energy-dependent effective mass, as described by Sercel et al.,[89,90] will help improving the accuracy of the model, making it more comparable with experimental results. Ghribi et al. recently used similar models, combined with exciton fine structure to study dielectric confinement in $CsPbBr_3$ NCs.[91] Our results on the dielectric tensor, band alignment and finite confinement effect of the NCs in $Cs_4PbBr_6$ and $Cspb_2Br_5$ could contribute to refining this model, explain experimental results and provide material design guidelines for Cs-Pb-Br composites. Additionally, this work could be extended to study other halide composites, such as $Cs_4PbCl_6$ and $Cs_4PbI_6$. These compounds have also been shown to coexist with their corresponding perovskite phases.[92–94] The change in the halide offers an additional degree of freedom (i.e., composition) for tuning the exciton emission in Cs-Pb-X (X=Br, Cl, I) composites.

## 3. Conclusions

In summary, we synthesized green-luminescent, ternary phases in the Cs-Pb-Br material system, namely perovskite-type $CsPbBr_3$ and non-perovskite-types $Cs_4PbBr_6$ and $CsPb_2Br_5$. We investigated their emission properties using PL experiments, which showed that the non-perovskite-type phases exhibit a higher quantum yield, with a blue shift of 50 meV and a red shift of 40 meV (with respect to the PL peak of bulk $CsPbBr_3$) for the $Cs_4PbBr_6$ and $CsPb_2Br_5$ phases, respectively. Microscopic characterization by means of correlative CL hyperspectral imaging, as well as EDX elemental mapping, showed that the luminescence emission at 2.45 eV and 2.36 eV of $Cs_4PbBr_6$ and $CsPb_2Br_5$ stems from nanocrystals of $CsPbBr_3$, embedded in the matrix material. We qualitatively explained the experimental results using a one-dimensional effective-mass model that considers electron-hole correlation, as well as band gap and dielectric mismatch at the interface of the nanocrystal and host materials. This model allowed us to rationalize the effect of the quantum and dielectric confinements on the emission shifts. We showed that including dielectric effects is important for estimating the emission energy and particle size. In the light of the correlated model, we conclude that the $CsPbBr_3@Cs_4PbBr_6$ composite is formed by small, strongly quantum-confined $CsPbBr_3$ nanocrystals in a $Cs_4PbBr_6$ matrix, while the $CsPbBr_3@CsPb_2Br_5$ composite is formed by large and weakly confined $CsPbBr_3$ nanocrystals in a $CsPb_2Br_5$ matrix, in which dielectric effects dominate, resulting in a net red shift.

# Acknowledgments

S.C.D. and D.A. are grateful for the financial support by the Helmholtz International Research School HI-SCORE (HIRS-0008). S.C.D. particularly thanks Prof. Dan Oron, Dr. Ayala Cohen and Dr. Anna Hirsch (Weizmann Institute of Science) for their support and valuable discussions on the optical model, as well as the DFPT calculations and analysis. M.S. further acknowledges the Deutsche Forschungsgemeinschaft (DFG, German Research Foundation) - project number 423749265 - SPP 2196 (SURPRISE), as well the Heisenberg program - project number 498155101 for funding. L.K. thanks the Minerva Centre for Self-Repairing Systems for Energy & Sustainability, the Aryeh and Mintzi Katzman Professorial Chair, and the Helen and Martin Kimmel Award for Innovative Investigation, for their support.

# References

[1] A. Kojima, K. Teshima, Y. Shirai, T. Miyasaka, *Journal of the American Chemical Society* **2009**, *131*, 6050.
[2] M. A. Green, E. D. Dunlop, J. Hohl-Ebinger, M. Yoshita, N. Kopidakis, X. Hao, *Progress in Photovoltaics: Research and Applications* **2021**, *29*, 657.
[3] S. D. Stranks, H. J. Snaith, *Nature Nanotechnology* **2015**, *10*, 391.
[4] T. M. Brenner, D. A. Egger, L. Kronik, G. Hodes, D. Cahen, *Nature Reviews Materials* **2016**, *1*, 15007.
[5] M. Ahmadi, T. Wu, B. Hu, *Advanced Materials* **2017**, *29*, 1.
[6] Q. Van Le, H. W. Jang, S. Y. Kim, *Small Methods* **2018**, *2*, 1700419.
[7] M. Kulbak, D. Cahen, G. Hodes, *Journal of Physical Chemistry Letters* **2015**, *6*, 2452.
[8] N. Yantara, S. Bhaumik, F. Yan, D. Sabba, H. A. Dewi, N. Mathews, P. P. Boix, H. V. Demir, S. Mhaisalkar, *J. Phys. Chem. Lett.* **2015**, *6*, 4360.
[9] X. Zhang, B. Xu, J. Zhang, Y. Gao, Y. Zheng, K. Wang, X. W. Sun, *Advanced Functional Materials* **2016**, *26*, 4595.
[10] C. Qin, T. Matsushima, A. S. D. Sandanayaka, Y. Tsuchiya, C. Adachi, *Journal of Physical Chemistry Letters* **2017**, *8*, 5415.
[11] Y. M. Chen, Y. Zhou, Q. Zhao, J. Y. Zhang, J. P. Ma, T. T. Xuan, S. Q. Guo, Z. J. Yong, J. Wang, Y. Kuroiwa, C. Moriyoshi, H. T. Sun, *ACS Applied Materials and Interfaces* **2018**, *10*, 15905.
[12] K. Lin, J. Xing, L. N. Quan, F. P. G. de Arquer, X. Gong, J. Lu, L. Xie, W. Zhao, D. Zhang, C. Yan, W. Li, X. Liu, Y. Lu, J. Kirman, E. H. Sargent, Q. Xiong, Z. Wei, *Nature* **2018**, *562*, 245.
[13] Y. F. Ng, N. F. Jamaludin, N. Yantara, M. Li, V. K. R. Irukuvarjula, H. V. Demir, T. C. Sum, S. Mhaisalkar, N. Mathews, *ACS Omega* **2017**, *2*, 2757.
[14] X. Li, Z. Wen, S. Ding, F. Fang, B. Xu, J. Sun, C. Liu, K. Wang, X. W. Sun, *Adv. Optical Mater.* **2020**, *8*, 2000232.
[15] X. Tang, Z. Hu, W. Yuan, W. Hu, H. Shao, D. Han, J. Zheng, J. Hao, Z. Zang, J. Du, Y. Leng, L. Fang, M. Zhou, *Advanced Optical Materials* **2017**, *5*, 1.
[16] N. Pourdavoud, T. Haeger, A. Mayer, P. J. Cegielski, A. L. Giesecke, R. Heiderhoff, S. Olthof, S. Zaefferer, I. Shutsko, A. Henkel, D. Becker-Koch, M. Stein, M. Cehovski, O. Charfi, H. Johannes, D. Rogalla, M. C. Lemme, M. Koch, Y. Vaynzof, K. Meerholz, W. Kowalsky, H. Scheer, P. Görrn, T. Riedl, *Advanced Materials* **2019**, *31*, 1903717.
[17] Y. Mao, C. Liang, G. Wang, Y. Wang, Z. Zhang, B. Wang, Z. Wen, Z. Mu, G. Sun, S. Chen, G. Xing, *Advanced Optical Materials* **2022**, *10*, 2201845.

[18] I. Dursun, M. De Bastiani, B. Turedi, B. Alamer, A. Shkurenko, J. Yin, A. M. El-Zohry, I. Gereige, A. AlSaggaf, O. F. Mohammed, M. Eddaoudi, O. M. Bakr, *ChemSusChem* **2017**, *10*, 3746.
[19] F. Palazon, C. Urso, L. De Trizio, Q. Akkerman, S. Marras, F. Locardi, I. Nelli, M. Ferretti, M. Prato, L. Manna, *ACS Energy Letters* **2017**, *2*, 2445.
[20] F. Palazon, S. Dogan, S. Marras, F. Locardi, I. Nelli, P. Rastogi, M. Ferretti, M. Prato, R. Krahne, L. Manna, *Journal of Physical Chemistry C* **2017**, *121*, 11956.
[21] Q. A. Akkerman, S. Park, E. Radicchi, F. Nunzi, E. Mosconi, F. De Angelis, R. Brescia, P. Rastogi, M. Prato, L. Manna, *Nano Letters* **2017**, *17*, 1924.
[22] M. Liu, J. Zhao, Z. Luo, Z. Sun, N. Pan, H. Ding, X. Wang, *Chemistry of Materials* **2018**, *30*, 5846.
[23] S. K. Balakrishnan, P. V. Kamat, *Chemistry of Materials* **2018**, *30*, 74.
[24] G. Maity, S. K. Pradhan, *Journal of Alloys and Compounds* **2020**, *816*, 152612.
[25] B. Turedi, K. J. Lee, I. Dursun, B. Alamer, Z. Wu, E. Alarousu, O. F. Mohammed, N. Cho, O. M. Bakr, *Journal of Physical Chemistry C* **2018**, *122*, 14128.
[26] S. Caicedo-Dávila, H. Funk, R. Lovrincic, C. Müller, C. Müller, M. Sendner, O. Cojocaru-Mirédin, F. Lehmann, R. Gunder, A. Franz, S. Levcenko, S. Levcenco, A. V. Cohen, L. Kronik, Benedikt Haas, B. Haas, C. Koch, D. Abou-Ras, *Journal of Physical Chemistry C* **2019**, *123*, 17666.
[27] Z.-L. Yu, Y.-Q. Zhao, Q. Wan, B. Liu, J.-L. Yang, M.-Q. Cai, *J. Phys. Chem. C* **2020**, *124*, 23052.
[28] M. Nikl, E. Mihokova, K. Nitsch, F. Somma, C. Giampaolo, G. P. Pazzi, P. Fabeni, S. Zazubovich, *Chemical Physics Letters* **1999**, *306*, 280.
[29] V. Babin, P. Fabeni, E. Mihokova, M. Nikl, G. P. Pazzi, *Physica Status Solidi B: Basic Research* **2000**, *205*, 205.
[30] C. D. Weerd, J. Lin, L. Gomez, Y. Fujiwara, K. Suenaga, T. Gregorkiewicz, *Journal of Physical Chemistry C* **2017**, *121*, 19490.
[31] L. N. Quan, R. Quintero-Bermudez, O. Voznyy, G. Walters, A. Jain, J. Z. Fan, X. Zheng, Z. Yang, E. H. Sargent, *Advanced Materials* **2017**, *29*, 1.
[32] X. Chen, F. Zhang, Y. Ge, L. Shi, S. Huang, J. Tang, Z. Lv, L. Zhang, B. Zou, H. Zhong, *Advanced Functional Materials* **2018**, *28*, 1.
[33] Z. Zhang, Y. Zhu, W. Wang, W. Zheng, R. Lin, F. Huang, *Journal of Materials Chemistry C* **2018**, *6*, 446.
[34] O. Nazarenko, M. R. Kotyrba, M. Wörle, E. Cuervo-Reyes, S. Yakunin, M. V. Kovalenko, *Inorganic Chemistry* **2017**, *56*, 11552.
[35] M. I. Saidaminov, J. Almutlaq, S. Sarmah, I. Dursun, A. A. Zhumekenov, R. Begum, J. Pan, N. Cho, O. F. Mohammed, O. M. Bakr, *ACS Energy Letters* **2016**, *1*, 840.
[36] J. Yin, Y. Zhang, A. Bruno, C. Soci, O. M. Bakr, J. L. Brédas, O. F. Mohammed, *ACS Energy Letters* **2017**, *2*, 2805.
[37] S. Seth, A. Samanta, *Journal of Physical Chemistry Letters* **2017**, *8*, 4461.
[38] M. De Bastiani, I. Dursun, Y. Zhang, B. A. Alshankiti, X. H. Miao, J. Yin, E. Yengel, E. Alarousu, B. Turedi, J. M. Almutlaq, M. I. Saidaminov, S. Mitra, I. Gereige, A. Alsaggaf, Y. Zhu, Y. Han, I. S. Roqan, J. L. Bredas, O. F. Mohammed, O. M. Bakr, *Chemistry of Materials* **2017**, *29*, 7108.
[39] J.-H. H. Cha, J. H. Han, W. Yin, C. Park, Y. Park, T. K. Ahn, J. H. Cho, D.-Y. Y. Jung, *The Journal of Physical Chemistry Letters* **2017**, *8*, 565.
[40] J. Yin, H. Yang, K. Song, A. M. El-Zohry, Y. Han, O. M. Bakr, J. L. Brédas, O. F. Mohammed, *Journal of Physical Chemistry Letters* **2018**, *9*, 5490.
[41] Y. K. Jung, J. Calbo, J. S. Park, L. D. Whalley, S. Kim, A. Walsh, *Journal of Materials Chemistry A* **2019**, *7*, 20254.
[42] T. Zhang, Z. Chen, Y. Shi, Q. H. Xu, *Nanoscale* **2019**, *11*, 3216.
[43] P. Acharyya, P. Pal, P. K. Samanta, A. Sarkar, S. K. Pati, K. Biswas, *Nanoscale* **2019**, *11*, 4025.

[44] K. Nitsch, V. Hamplová, M. Nikl, K. Polák, M. Rodová, *Chemical Physics Letters* **1996**, *258*, 518.

[45] Y. Rakita, N. Kedem, S. Gupta, A. Sadhanala, V. Kalchenko, M. L. Böhm, M. Kulbak, R. H. Friend, D. Cahen, G. Hodes, *Crystal Growth and Design* **2016**, *16*, 5717.

[46] Q. A. Akkerman, S. G. Motti, A. R. Srimath Kandada, E. Mosconi, V. D'Innocenzo, G. Bertoni, S. Marras, B. A. Kamino, L. Miranda, F. De Angelis, A. Petrozza, M. Prato, L. Manna, A. Ram, S. Kandada, V. D. Innocenzo, G. Bertoni, S. Marras, B. A. Kamino, F. D. Angelis, A. Petrozza, M. Prato, L. Manna, Q. A. Akkerman, S. G. Motti, A. Ram, S. Kandada, E. M. Σ, *Journal of the American Chemical Society* **2016**, *138*, 1010.

[47] J. Liang, J. Liu, Z. Jin, *Solar RRL* **2017**, *1*, 1770138.

[48] J. Lei, F. Gao, H. Wang, J. Li, J. Jiang, X. Wu, R. Gao, Z. Yang, S. (Frank) Liu, *Solar Energy Materials and Solar Cells* **2018**, *187*, 1.

[49] Z. Qin, S. Dai, V. G. Hadjiev, C. Wang, L. Xie, Y. Ni, C. Wu, G. Yang, S. Chen, L. Deng, Q. Yu, G. Feng, Z. M. Wang, J. Bao, *Chemistry of Materials* **2019**.

[50] N. Riesen, M. Lockrey, K. Badek, H. Riesen, *Nanoscale* **2019**, *11*, 4001.

[51] M. Shin, S.-W. Nam, A. Sadhanala, R. Shivanna, M. Anaya, A. Jiménez-Solano, H. Yoon, S. Jeon, S. D. Stranks, R. L. Z. Hoye, B. Shin, *ACS Appl. Energy Mater.* **2020**, *3*, 192.

[52] Z. Ma, F. Li, D. Zhao, G. Xiao, B. Zou, *CCS Chemistry* **2020**, *2*, 71.

[53] S. Aharon, L. Etgar, *Nano Select* **2021**, *2*, 83.

[54] W. Castro Ferreira, B. S. Araújo, M. A. P. Gómez, F. E. O. Medeiros, C. W. de Araujo Paschoal, C. B. da Silva, P. T. C. Freire, U. F. Kaneko, F. M. Ardito, N. M. Souza-Neto, A. P. Ayala, *J. Phys. Chem. C* **2022**, *126*, 541.

[55] J. Xu, W. Huang, P. Li, D. R. Onken, C. Dun, Y. Guo, K. B. Ucer, C. Lu, H. Wang, S. M. Geyer, R. T. Williams, D. L. Carroll, *Advanced Materials* **2017**, *29*, 1.

[56] Q. A. Akkerman, A. L. Abdelhady, L. Manna, *Journal of Physical Chemistry Letters* **2018**, *9*, 2326.

[57] U. Petralanda, G. Biffi, S. C. Boehme, D. Baranov, R. Krahne, L. Manna, I. Infante, *Nano Letters* **2021**, *21*, 8619.

[58] L. E. Brus, *The Journal of Chemical Physics* **1984**, *80*, 4403.

[59] L. Brus, *J. Phys. Chem.* **1986**, *90*, 2555.

[60] S. Caicedo-Dávila, R. R. Gunder, J. A. Márquez, S. Levcenco, K. Schwarzburg, T. Unold, D. Abou-Ras, S. Caicedo-Davila, R. R. Gunder, J. A. Marquez, S. Levcenko, K. Schwarzburg, T. Unold, D. Abou-Ras, *The Journal of Physical Chemistry C* **2020**, *124*, 19514.

[61] T. Schmidt, K. Lischka, W. Zulehner, *Physical Review B* **1992**, *45*, 8989.

[62] A. Zubiaga, J. A. García, F. Plazaola, V. Muñoz-Sanjosé, C. Martínez-Tomás, *Phys. Rev. B* **2003**, *68*, 245202.

[63] C. Spindler, T. Galvani, L. Wirtz, G. Rey, S. Siebentritt, *Journal of Applied Physics* **2019**, *126*.

[64] Y. Kajino, S. Otake, T. Yamada, K. Kojima, T. Nakamura, A. Wakamiya, Y. Kanemitsu, Y. Yamada, *Phys. Rev. Materials* **2022**, *6*, L043001.

[65] P. Villars, K. Cenzual, *$Cs_4PbBr_6$ Crystal Structure: Datasheet from "PAULING FILE Multinaries Edition – 2012" in SpringerMaterials*, Springer-Verlag Berlin Heidelberg & Material Phases Data System (MPDS), Switzerland & National Institute for Materials Science (NIMS), Japan.

[66] Chr. K. Moller, *Nature* **1958**, *182*, 1436.

[67] P. Villars, K. Cenzual, Eds., *$CsPb_2Br_5$ (T = 300 K) crystal structure: Datasheet from "PAULING FILE multinaries edition – 2012" in SpringerMaterials (https://materials.springer.com/isp/crystallographic/docs/sd_1533100)*, Springer-Verlag Berlin Heidelberg & Material Phases Data System (MPDS), Switzerland & National Institute for Materials Science (NIMS), Japan.

[68] L. Bányai, S. W. Koch, *Semiconductor Quantum Dots*, Vol. Volume 2, World Scientific, **1993**.

[69] D. Fröhlich, K. Heidrich, H. Künzel, G. Trendel, J. Treusch, *Journal of Luminescence* **1979**, *18–19*, 385.

[70] M. Sebastian, J. A. Peters, C. C. Stoumpos, J. Im, S. S. Kostina, Z. Liu, M. G. Kanatzidis, A. J. Freeman, B. W. Wessels, *Physical Review B - Condensed Matter and Materials Physics* **2015**, *92*, 1.
[71] L. Protesescu, S. Yakunin, M. I. Bodnarchuk, F. Krieg, R. Caputo, C. H. Hendon, R. X. Yang, A. Walsh, M. V. Kovalenko, *Nano Letters* **2015**, *15*, 3692.
[72] G. R. Yettapu, D. Talukdar, S. Sarkar, A. Swarnkar, A. Nag, P. Ghosh, P. Mandal, *Nano Letters* **2016**, *16*, 4838.
[73] M. Kumagai, T. Takagahara, *Physical Review B* **1989**, *40*, 12359.
[74] P. G. Bolcatto, C. R. Proetto, *Physical Review B - Condensed Matter and Materials Physics* **1999**, *59*, 12487.
[75] J. L. Movilla, J. Planelles, *Computer Physics Communications* **2005**, *170*, 144.
[76] J. Planelles, *Theoretical Chemistry Accounts* **2017**, *136*, 1.
[77] F. Rajadell, J. I. Climente, J. Planelles, *Physical Review B* **2017**, *96*, 1.
[78] L. V. Keldysh, *Pis'ma Zh. Eksp. Teor. Fiz.* **1979**, *29*, 716.
[79] T. Takagahara, *Physical Review B* **1993**, *47*, 4569.
[80] P. G. Bolcatto, C. R. Proetto, *Journal of Physics Condensed Matter* **2001**, *13*, 319.
[81] F. Rajadell, J. L. Movilla, M. Royo, J. Planelles, *Physical Review B - Condensed Matter and Materials Physics* **2007**, *76*, 1.
[82] B. Kang, K. Biswas, *J. Phys. Chem. Lett.* **2018**, *9*, 830.
[83] G. Hu, W. Qin, M. Liu, X. Ren, X. Wu, L. Yang, S. Yin, *J. Mater. Chem. C* **2019**, *7*, 4733.
[84] Q. Zhang, X. Sun, W. Zheng, Q. Wan, M. Liu, X. Liao, T. Hagio, R. Ichino, L. Kong, H. Wang, L. Li, *Chem. Mater.* **2021**, *33*, 3575.
[85] Z.-P. Huang, B. Ma, H. Wang, N. Li, R.-T. Liu, Z.-Q. Zhang, X.-D. Zhang, J.-H. Zhao, P.-Z. Zheng, Q. Wang, H.-L. Zhang, *J. Phys. Chem. Lett.* **2020**, *11*, 6007.
[86] L. Ding, B. Borjigin, Y. Li, X. Yang, X. Wang, H. Li, *ACS Appl. Mater. Interfaces* **2021**, *13*, 51161.
[87] S. Cho, S. H. Yun, *Commun Chem* **2020**, *3*, 15.
[88] S. Datta, *Quantum Transport: Atom to Transistor*, Cambridge University Press, **2005**.
[89] P. C. Sercel, J. L. Lyons, N. Bernstein, A. L. Efros, *J. Chem. Phys.* **2019**, *151*, 234106.
[90] P. C. Sercel, J. L. Lyons, D. Wickramaratne, R. Vaxenburg, N. Bernstein, A. L. Efros, *Nano Letters* **2019**, *19*, 4068.
[91] A. Ghribi, R. Ben Aich, K. Boujdaria, T. Barisien, L. Legrand, M. Chamarro, C. Testelin, *Nanomaterials* **2021**, *11*, 3054.
[92] M. Nikl, E. Mihokova, K. Nitsch, *Solid State Communications* **1992**, *84*, 1089.
[93] S. Kondo, K. Amaya, S. Higuchi, T. Saito, H. Asada, M. Ishikane, *Solid State Communications* **2001**, *120*, 141.
[94] S. Kondo, A. Masaki, T. Saito, H. Asada, *Solid State Communications* **2002**, *124*, 211.

# Supporting Information for: Effects of Quantum and Dielectric Confinement on the Emission of Cs-Pb-Br Composites

Sebastián Caicedo-Dávila[1,b,*], Pietro Caprioglio[2,3], Frederike Lehmann[1], Sergiu Levcenco[1], Martin Stolterfoht[2], Dieter Neher[2], Leeor Kronik,[4] and Daniel Abou-Ras[1]

[1]Helmholtz-Zentrum Berlin für Materialien und Energien, Hahn-Meitner-Platz 1. 14109 Berlin, Germany
[2]Institute of Physics and Astronomy, University of Potsdam, 14476 Potsdam, Germany
[3]Department of Physics, University of Oxford, Clarendon Laboratory, Parks Road, Oxford, UK
[4]Department of Molecular Chemistry and Materials Science, Weizmann Institute of Science, Rehovoth 76100, Israel
*sebastian.caicedo@tum.de

## S.1. Synthesis of the Samples

$CsPbBr_3$ was synthesized from an equimolar mixture of CsBr (99.99% from Ossila) and $PbBr_2$ (98%, extra pure, from Arcos Organics) in dimethylformamide (DMF, 99.8%, Roth) and left to stir overnight at 60 °C followed by evaporation of the solvent at 85 °C and annealing at 140 °C for 1 h. The resulting material is a fine, bright-orange powder (see insets in Figure 1 in the main text). For the synthesis of $CsPb_2Br_5$ a 1:3 ($CsBr:PbBr_2$) molar ratio solution was used, followed by the same synthesis procedure as for $CsPbBr_3$. The resulting material is a white powder.

$Cs_4PbBr_6$ crystals were synthesized using the anti-solvent method, adapted from the one reported by Cha et al.[1] A mixture of 0.2554 g (0.2 $\mathcal{M}$) of CsBr (99.99% from Ossila) and 0.1079 g (0.05 $\mathcal{M}$) $PbBr_2$ (98%, extra pure, from Arcos Organics) in dimethyl sulfoxide (≥ 99.99%, Rotisolv, Roth) was dissolved under ultrasonication ($T < 40$ °C) until no precipitation was observed. The solution was filtered (0.2 μm pore-size syringe filter) and transferred into a 50 ml glass vial and sealed with the cap, which was perforated by a needle to ensure diffusion of the solvent through a small hole. The vials were place into a 250 ml solvent vial with screw cap under dichloromethane (DCM, ≥99.99%, Rotisolv, Roth) atmosphere. Within few days, yellow crystal seeds were removed from the vials, washed with DMF and DCM solution. The crystals are highly green luminescent under blue illumination (Figure 1 of the main text). The samples were stored under $N_2$ to avoid potential degradation by oxygen and humidity.

## S.2. Characterization of the Samples

**X-ray diffraction** patterns were measured using a PANalytical X'Pert MDP Pro diffractometer at room temperature. A Cu-Kα radiation source with λ=0.15406 nm was used in Bragg-Brentanon configuration. The powder diffractograms for $CsPbBr_3$ and $CsPb_2Br_5$ are shown in Figure S1. Due to the small size and amount of $Cs_4PbBr_6$ crystals, it was not possible to perform the measurement. However, our EDX compositional analysis showed almost identical composition of the crystals to that obtained in a thin film, as detailed in a previous work.[2]

---

[b] Present Address: Physics Department, TUM School of Natural Sciences
Technical University of Munich, 85748 Garching

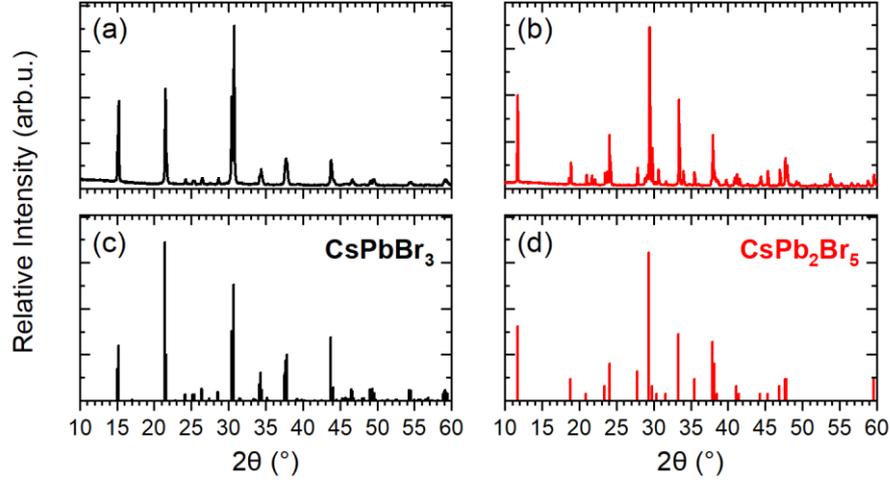

Figure S1. XRD patterns of (a) CsPbBr$_3$ and (b) CsPb$_2$Br$_5$ powders synthesized as described in Section S.1. Refernce intensities for (c) CsPbBr$_3$ PDF 01-072-7929 (Ref [3]) and (b) CsPb$_2$Br$_5$ PDF 00-025-0211 (Ref [4]) are also shown and used for phase identification.

**Steady-state PL** spectra in Figure 1 were measured at room temperature using a 409 nm diode laser, under an excitation power density of ~0.28 W/cm$^2$ and a laser spot size of ~100 μm. The PL emission was dispersed by a 0.5 m grating monochromator and detected by a charge-coupled device (CCD) camera. Light in the spectral range from 1.7 to 2.8 eV was detected with a spectral resolution of 5 meV

**Cathodoluminescence** panchromatic and band-pass maps – filtered at 500 nm, bandwidth 50 nm – intensity maps were acquired using a SPARC system from Delmic in a Zeiss MERLIN scanning electron microscope (SEM), at acceleration voltages of up to 5 kV (see text for specific measurement conditions) and a beam current of 50 pA to ensure no sample damage. Hyperspectral maps were acquired with a Kymera 193i spectrograph from Andor. Light was focused onto a monochromator grating of 300 lines/mm blazed at 500 nm. The exposure/dwelling time was adjusted below 50 ms, ensuring good signal-to-noise ratio in each measurement. The background was measured using the same experimental conditions with the electron beam blanked and it was subtracted from the CL maps. Furthermore, hyperspectral maps were also compensated by the instrument spectral response.

The practical resolution is determined by the experimental parameters (electron beam) considering the Gruen's penetration depth,[5] in the approximation for low-acceleration voltages proposed by Everhart and Hoff:[6]

$$R_e = \left(\frac{0.0398}{\rho}\right) E_b^{1.75} \text{ [μm]} \tag{S1}$$

where $\rho$ is the material density in g/cm$^3$ and $E_b$ is the electron beam energy in keV.

***Energy-dispersive X-ray Spectroscopy*** (EDX) maps were measured at an acceleration voltage of 5 kV. The X-ray fluorescence was measured using a windowless Oxford Instruments Ultim Extreme X-ray detector with enhanced sensitivity to low-energy X-ray lines, which facilitates measurements under low voltage conditions. The Cs-M, Pb-M and Br-L X-ray characteristic lines were used for element identification. One major drawback of these measurement conditions is that they hinder the quantification of the composition, owing to the low acceleration voltage. Furthermore, the Cs-M line lies closely to the O-K line, which complicates the deconvolution of the elements for proper quantification. Therefore, Figures 2f and 3 show elemental maps and normalized line-scans, which allow us to extract changes in the relative composition among phases.

## S.3. Effective Mass Exciton Model Including Electron-Hole Correlation

The exciton in a $CsPbBr_3$ NC, confined in a host matrix of $Cs_4PbBr_6$ or $CsPb_2Br_5$, is modelled using a nanoplatelet effectively confined only in the $z$ direction, with the perpendicular $x, y$ plane modeled as a square with length $L = 30$ nm, as proposed by Rajadell and Planelles.[7] This approximation allows us to understand how the dielectric confinement depends on the NC size and interpret our experimental results. However, it does not offer quantitative information. The model approximates the effects of the self-interaction and generalized Coulomb interaction, arising from the dielectric mismatch at the interface, using the method of image charges as described by Kumagai.[8]

The effective mass, exciton Hamiltonian, and exciton wave function are given in Eqs. (1)-(4) in the main text. Here, we present the single-particle equations for completeness. We start with the single-particle Hamiltonian to describe either the electron or the hole in the system:

$$H^{(i)}(\boldsymbol{r}_i) = -\nabla_i \frac{h^2}{2m_i^*} \nabla_i + V_{\text{pot}}(\boldsymbol{r}_i) + V_i^{\text{self}}(\boldsymbol{r}_i) \quad (S2)$$

where the first term is the kinetic energy operator, $V_{\text{pot}}$ is the confining potential, which arises from band discontinuity at the interface between the NCs and the host material, and $V_i^{\text{self}}$ is the self-interaction potential, i.e., the interaction of a charge with the induced charges, caused by the dielectric mismatch at the interface. This self-interaction potential is calculated using the method of image charges in a 1D quantum well:[7,8]

$$V_i^{\text{self}}(r_i) = \sum_{n=\pm 1, \pm 2, \ldots} \frac{q_n e^2}{2\varepsilon_1 [z_i - (-1)^n z_i + nL]} \quad (S3)$$

where $\varepsilon_1$ and $\varepsilon_2$ are the dielectric constants of the confined NC and the host material, respectively, $z_i$ is the particle coordinate in the confined direction, $L$ is the width of the well (NC size) and $q_n = \left(\frac{\varepsilon_1 - \varepsilon_2}{\varepsilon_1 + \varepsilon_2}\right)^{|n|}$. The wave function is found by minimizing the exciton emission energy $E_{\text{em}} = \langle \Psi(\boldsymbol{r}_e, \boldsymbol{r}_h) | H(\boldsymbol{r}_e, \boldsymbol{r}_h) | \Psi(\boldsymbol{r}_e, \boldsymbol{r}_h) \rangle$. Details of the calculation, as well as the Mathematica code used for our calculations, can be found in references [7,9].

## S.4. Calculation of the High-frequency Dielectric Constants

Calculations of high-frequency dielectric constants were performed using the Vienna Ab-initio Simulation Package (VASP),[10] based on the Perdew-Burke-Ernzerhof (PBE) form of the generalized gradient approximation to describe the exchange-correlation interactions[11] and including spin-orbit coupling. Dispersion interactions were included within the Tkatchenko-Scheffler scheme,[12] using an iterative Hirshfeld partitioning of the charge-density.[13,14] Core electrons were described by the projector augmented waves method.[15,16] The orthorhombic structure of $CsPbBr_3$[3] was used for the electronic structure calculations. A planewave basis set with cutoff of 700 eV was used and $k$-space integration was carried out on a $6 \times 6 \times 6$ grid. For $Cs_4PbBr_6$, the atomic structure reported by Velázquez et al.[17] was used as a starting point. A cutoff of 500 eV and a $4 \times 4 \times 3$ $k$-space grid was used. The structure of $CsPb_2Br_5$ reported by Cola and Riccardi[4] was used as a starting point; a cutoff of 700 eV and a $3 \times 3 \times 2$ $k$-space grid were used. The geometries were optimized using a conjugated gradient algorithm until the largest force was below 0.001 eV/Å. The high-frequency dielectric constants were calculated using density functional perturbation theory as implemented by Gajdoš et al.[18] in the VASP package. Convergence was tested for all the parameters in the calculations.

Figure S1 summarizes the calculated dielectric tensors of all of the Cs-Pb-Br ternary phases. A particular feature of the dielectric tensors of CsPbBr$_3$ and CsPb$_2$Br$_5$ is their anisotropy. For CsPbBr$_3$, the dielectric constant is larger along the longest lattice vector of the orthorhombic structure ($b = 11.75$ Å). The second largest dielectric constant is along the $a = 8.54$ Å lattice vector, and the smallest one is along the $c = 8.06$ Å direction. This suggests that in the CsPbBr$_3$ the polarizability increases as the unit cell is stretched, which results from the tilting of the PbBr$_6$ octahedra. This phenomenon appears to be related to the one described by Kang and Biswas,[19] in which the cross-gap hybridization of Pb–Br p-orbitals results in large Born effective charges and dielectric constants. The dielectric constant of Cs$_4$PbBr$_6$ is isotropic, which can be understood when looking at the primitive cell in Figure S1b. The disjoint octahedra not only reduce the band dispersion, but also hinder the lattice polarization, which results in a lower dielectric constant. The orientation of the octahedra relative to one another seems to have little influence on the dielectric response of the material. Finally, CsPb$_2$Br$_5$ exhibits strong polarizability on the $ab$ plane, owing to the face-sharing PbBr$_2$ prisms forming planes (Figure S1c). This results in an in-plane dielectric constant that is even larger than that of CsPbBr$_3$, regardless of the larger band-gap energy. The polarizability in the $c$ direction is closer to that of Cs$_4$PbBr$_6$. The discontinuous planes inhibit lattice polarization and reduce band dispersion, which in turn reduces the dielectric constant.

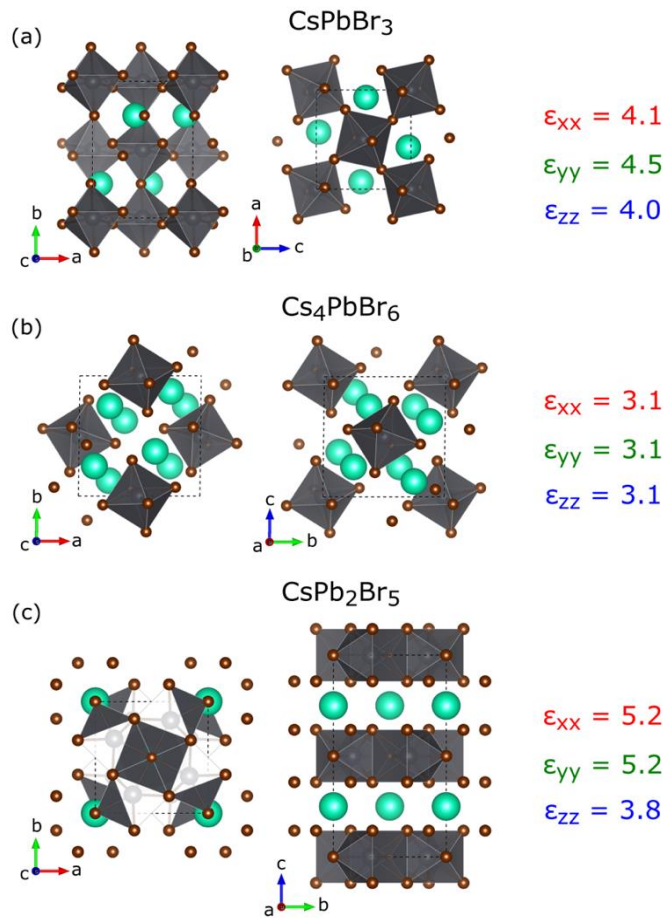

**Figure S2.** Schematic representations of (a) CsPbBr$_3$, (b) Cs$_4$PbBr$_6$, and (c) CsPb$_2$Br$_5$. The diagonal elements of the dielectric tensor are color-coded to the lattice vector. The geometry of the crystal determines the magnitude of the dielectric constant (see text for details).

In the calculations of the exciton model, we use a dielectric constant of $\varepsilon_1 = 4.2$ for the confined CsPbBr$_3$ (average of $\varepsilon$ in all the directions, because the variation is small). The dielectric constant of the host materials is $\varepsilon_2 = 3.1$ or $3.8$ for Cs$_4$PbBr$_6$ and CsPb$_2$Br$_5$, respectively. The lowest $\varepsilon$ of the dielectric tensor of $CsPb_2Br_5$ is used, because the effect of the dielectric mismatch is strongest when $\varepsilon_1 > \varepsilon_2$. Naturally, the anisotropy of $CsPb_2Br_5$ will reduce the dielectric confinement in two directions.

Table S1. Electron and hole effective masses $m^*_{e/h}$ (Ref. [20]), dielectric constants $\varepsilon$, band-gap energies $E_g$ and the band mismatch parameter $\Delta E_g/2$, used in the model of the finite potential barriers

|  | $CsPbBr_3$ | $Cs_4PbBr_6$ | $CsPb_2Br_5$ |
|---|---|---|---|
| $m^*_e$ | 0.15 | | |
| $m^*_h$ | 0.14 | | |
| $\varepsilon$ | 4.2 | 3.1 | 3.8 |
| $E_g$ | 2.3 | 4.0 | 3.7 |
| $\Delta E_g/2$ | – | 1.7 | 1.4 |

# References

[1] J.-H. H. Cha, J. H. Han, W. Yin, C. Park, Y. Park, T. K. Ahn, J. H. Cho, D.-Y. Y. Jung, *The Journal of Physical Chemistry Letters* **2017**, *8*, 565.
[2] S. Caicedo-Dávila, H. Funk, R. Lovrincic, C. Müller, C. Müller, M. Sendner, O. Cojocaru-Mirédin, F. Lehmann, R. Gunder, A. Franz, S. Levcenko, S. Levcenco, A. V. Cohen, L. Kronik, Benedikt Haas, B. Haas, C. Koch, D. Abou-Ras, *Journal of Physical Chemistry C* **2019**, *123*, 17666.
[3] M. Rodová, J. Brožek, K. Nitsch, *Journal of thermal analysis* **2003**, *71*, 667.
[4] M. Cola, R. Riccardi, *Zeitschrift für Naturforschung A* **1971**, *26*, 1328.
[5] B. G. Yacobi, D. B. Holt, *Cathodoluminescence Microscopy of Inorganic Solids*, **1990**.
[6] T. E. Everhart, P. H. Hoff, *Journal of Applied Physics* **1971**, *42*, 5837.
[7] F. Rajadell, J. I. Climente, J. Planelles, *Physical Review B* **2017**, *96*, 1.
[8] M. Kumagai, T. Takagahara, *Physical Review B* **1989**, *40*, 12359.
[9] J. Planelles, *Theoretical Chemistry Accounts* **2017**, *136*, 1.
[10] G. Kresse, J. Furthmüller, *Phys. Rev. B* **1996**, *54*, 11169.
[11] J. P. Perdew, K. Burke, Y. Wang, *Physical Review B* **1996**, *54*, 16533.
[12] A. Tkatchenko, M. Scheffler, *Phys. Rev. Lett.* **2009**, *102*, 073005.
[13] T. Bučko, S. Lebègue, J. Hafner, J. G. Ángyán, *Journal of Chemical Theory and Computation* **2013**, *9*, 4293.
[14] T. Bučko, S. Lebègue, J. G. Ángyán, J. Hafner, *Journal of Chemical Physics* **2014**, *141*, 034114.
[15] P. E. Blöchl, *Physical Review B* **1994**, *50*, 17953.
[16] G. Kresse, D. Joubert, *Phys. Rev. B* **1999**, *59*, 1758.
[17] M. Velázquez, A. Ferrier, S. Péchev, P. Gravereau, J. P. Chaminade, X. Portier, R. Moncorgé, *Journal of Crystal Growth* **2008**, *310*, 5458.
[18] M. Gajdoš, K. Hummer, G. Kresse, J. Furthmüller, F. Bechstedt, *Phys. Rev. B* **2006**, *73*, 1.
[19] B. Kang, K. Biswas, *J. Phys. Chem. Lett.* **2018**, *9*, 830.
[20] L. Protesescu, S. Yakunin, M. I. Bodnarchuk, F. Krieg, R. Caputo, C. H. Hendon, R. X. Yang, A. Walsh, M. V. Kovalenko, *Nano Letters* **2015**, *15*, 3692.